\documentclass[nofootinbib, aps,twocolumn,showpacs,preprintnumbers,amssymb,superscriptaddress,10pt]{revtex4-1}

\usepackage{amsmath,amssymb,amsfonts,graphicx,epsfig,todonotes}
\usepackage[noconfig]{refstyle}
\usepackage[hyperfootnotes=true, linktocpage=true, citecolor=blue, linkcolor=blue, urlcolor=Maroon, filecolor=Maroon]{hyperref}

\let\eqref=\relax
\newref{eq}{name={},Name={Eq.~},names={eqs.~},Names={Eqs.~},rngtxt={-},refcmd=(\ref{#1})}
\newref{tab}{name={Table~},Name={Table~},names={tables~},Names={Tables~}}
\newref{sec}{name={Section~},Name={Section~},names={sections~},Names={Sections~}}
\newref{fig}{name={Figure~},Name={Figure~},names={figures~},Names={Figures~}}
\numberwithin{equation}{section}

\newcommand{\eeq}{\end{equation}}
\newcommand{\beq}{\begin{equation}}

\begin{document}

\begin{flushright} 
 {~\\CERN-TH-2020-050 \\ CTPU-PTC-20-06} 
\end{flushright}

\title{Machine Learning String Standard Models}

\author{Rehan Deen}
\affiliation{Rudolf Peierls Centre for Theoretical Physics, University of Oxford, \\Parks Road, Oxford OX1 3PU, UK} 
\email[]{rehan.deen@physics.ox.ac.uk\\ andre.lukas@physics.ox.ac.uk}

\author{Yang-Hui He}
\affiliation{School of Physics, NanKai University, Tianjin, 300071, P.R. China}
\affiliation{Department of Mathematics, City, University of London, London EC1V0HB, UK}
\affiliation{Merton College, University of Oxford, OX1 4JD, UK}
\email[]{hey@maths.ox.ac.uk}

\author{Seung-Joo Lee}
\affiliation{Center for Theoretical Physics of the Universe, Institute for Basic Science, Daejeon 34051, South Korea}
\affiliation{CERN, Theory Department, 1 Esplande des Particules, Geneva 23, CH-1211, Switzerland}
\email[]{seung.joo.lee@cern.ch}

\author{Andre Lukas}
\affiliation{Rudolf Peierls Centre for Theoretical Physics, University of Oxford, \\Parks Road, Oxford OX1 3PU, UK}
\email[]{andre.lukas@physics.ox.ac.uk}


\begin{abstract}\noindent
We study machine learning of phenomenologically relevant properties of string compactifications, which arise in the context of heterotic line bundle models. Both supervised and unsupervised learning are considered. We find that, for a fixed compactification manifold, relatively small neural networks are capable of distinguishing consistent line bundle models with the correct gauge group and the correct chiral asymmetry from random models without these properties. The same distinction can also be achieved in the context of unsupervised learning, using an auto-encoder. Learning non-topological properties, specifically the number of Higgs multiplets, turns out to be more difficult, but is possible using sizeable networks and feature-enhanced data sets.
\end{abstract}



\pacs{}

\maketitle

\section{Introduction}
Techniques from machine learning have recently been introduced into string theory~\cite{He:2017aed,He:2017set,Krefl:2017yox,Carifio:2017bov,Ruehle:2017mzq} and have been explored for a variety of different machine learning architectures and string theory settings (for reviews see Refs.~\cite{Ruehle:2020jrk,He:2018jtw} and references therein). One of the main motivations for bringing modern data science methods to string theory is the vast amount of data generated by string theory. String theory data describes different solutions to the theory and, at the most basic level of topological or quasi-topological properties, is discrete data described by sets of integers, which specifies the compactification manifold and branes/fluxes thereon.  

For a given string solution this integer data determines (to a large extent) the massless spectrum of the associated low-energy theory. It is widely believed that the set of (topological types of) string solutions is finite, although no general proof is known. (For instance, it is a conjecture of Yau that the topological types of compact smooth connected Calabi-Yau manifolds is finite in every dimension.) If this is indeed the case it is clear, however, that the set is vast and an impressive lower bound of $\sim 10^{272000}$ has been given in Ref.~\cite{Taylor:2015xtz}. Can machine learning help to deal with this vast set of string vacua, for example by organising the data or distinguishing between models with different properties?

One of the central problems of string theory remains understanding its relation to established low-energy particle physics. It is now known that string theory contains many models with a promising spectrum of particles and, in the context of a specific construction, a lower bound of $\sim 10^{23}$ (``a mole of models"~\cite{mole}) for the number of such models has been given in Ref.~\cite{Constantin:2018xkj}. Nevertheless, string solutions which lead to such models constitute a very small fraction of the total. Finding phenomenologically promising models within string theory and even deciding whether given string vacua are promising can be non-trivial tasks. In this paper we will deal with the second problem and the main question we will address is whether machine learning can distinguish string vacua which lead to phenomenologically attractive models from those which do not. In other words, can a neural network learn whether or not a string solution is of physical interest?

We will be addressing this question in the context of heterotic line bundle models~\cite{Anderson:2011ns,Anderson:2012yf,Anderson:2013xka}, a class of models which has the virtue of being conceptually relatively simple and for which sizeable sets of phenomenologically  promising models are known. This means that training sets for machine learning can readily be constructed. To be clear about terminology, by ``standard-like" model (SLM for short) we mean a consistent string solution which gives rise to the standard model gauge group and has the correct chiral asymmetry of three families of quarks and leptons. Whether a string solution is a SLM is a topological question, that is, it only depends on topological quantities such as Chern classes and indices. Further requirements on a SLM towards realistic physics are the absence of vector-like matter and the presence of Higgs doublets. These properties are controlled by bundle cohomology and are less robust. They can depend on the (complex structure) moduli of the model and, in this sense, are non-topological.  

For the purpose of this paper, we will only consider supervised learning and unsupervised learning with auto-encoders, either based on fully-connected feed-forward networks. Unsupervised learning for heterotic orbifold models with auto-encoders has been studied in Ref.~\cite{Mutter:2018sra} and reinforcement learning of string models has been investigated in Refs.~\cite{Halverson:2019tkf,Larfors:2020ugo}.

In the next section, we briefly review heterotic line bundle models and in \secref{data} we describe the associated data sets. In \secref{smsimple} we show that simple fully connected networks with supervised learning can distinguish SLMs and non SLMs. This can also be done via unsupervised learning, using an auto-encoder, as explained in \secref{smauto}. Finally, in \secref{Higgs}, we investigate whether neural networks can detect the presence or absence of Higgs multiplets in SLMs. We find this is possible, but requires a more careful approach which draws on previous experience with line bundle cohomology~\cite{Constantin:2018hvl,Klaewer:2018sfl,Larfors:2019sie,Brodie:2019pnz} and machine learning of line bundle cohomology formulae~\cite{Brodie:2019dfx}. We conclude in \secref{conclusion}.

\section{Heterotic line bundle models}\seclabel{lbmodels}
Heterotic line bundle models have been introduced and analysed in Refs.~\cite{Anderson:2011ns,Anderson:2012yf,Anderson:2013xka} and we refer to these papers for details. Here, we present a concise summary with emphasis on aspects relevant to machine learning applications.
The data which specifies a heterotic line bundle model is a tuple $(X,\Gamma, V)$, where $X$ is a Calabi-Yau three-fold, $\Gamma$ is a freely-acting symmetry on $X$ and $V$ is a ($\Gamma$-equivariant) line bundle sum on $X$, here taken to be of rank five\footnote{Line bundle sums with different rank, specifically with rank four, can also lead to phenomenologically promising models. However, Wilson line breaking of such models requires large symmetries $\Gamma$ which are rare. As a result, large data sets are not yet available.}
\begin{equation}
 V=\bigoplus_{a=1}^5 L_a\; , \eqlabel{V}
\end{equation} 
where $L_a\rightarrow X$ are line bundles. For this to define a consistent string compactification on $X$ we require that
\begin{align}
\nonumber
 &c_1(V)=\sum_{a=1}^5c_1(L_a)\stackrel{!}{=}0\; ,\\ 
 &c_2(TX)-c_2(V)\in \mbox{Mori cone of }X\; , \eqlabel{ccond}
\end{align}
where the first condition is required for an embedding of the structure group of $V$ into $E_8$ and the second condition guarantees that the anomaly cancelation condition can be satisfied. A further consistency condition, which serves to ensure that the bundle $V$ preserves supersymmetry, is that the slopes
\begin{equation}
 \mu(L_a):=\int_Xc_1(L_a)\wedge J^2\stackrel{!}{=}0 \eqlabel{slopecond}
\end{equation}
for all five line bundles vanish simultaneously for some K\"ahler class $J$ of $X$. 

Such a consistent model on $X$ defines a four-dimensional $N=1$ grand unified theory (GUT) with gauge group $SU(5)$ and matter fields in the $SU(5)$ multiplets ${\bf 10}$, $\overline{\bf 10}$, $\bar{\bf 5}$ and ${\bf 5}$ (plus fields uncharged under $SU(5)$).\footnote{Additional Green-Schwarz anomalous $U(1)$ gauge symmetries are also present but their associated vector bosons are usually supermassive.} We recall that one standard model family fits precisely into the $SU(5)$ multiplet ${\bf 10}\oplus\bar{\bf 5}$. The numbers of these multiplets are governed by the bundle cohomologies
\begin{align}
\nonumber
 \#{\bf 10}=h^1(X,V)\;,\quad \#\overline{\bf 10}=h^2(X,V)\;,\\
\#\bar{\bf 5}=h^1(X,\wedge^2V)\;,\quad \#{\bf 5}=h^2(X,\wedge^2 V)\; .
\end{align} 
The chiral asymmetry is measured by the index of the bundle $V$ and is given by\footnote{This follows from $h^0(X,L)=h^3(X,L)=0$ which holds for line bundles $L$ with vanishing slope and from ${\rm ind}(V)={\rm ind}(\wedge^2 V)$, a general property of rank five bundles $V$.}
\begin{align}
\nonumber
 \#\mbox{GUT families} &= \#{\bf 10}-\#\overline{\bf 10}=\#\bar{\bf 5}-\#{\bf 5}\\
 &=-{\rm ind}(V)\stackrel{!}{=}3|\Gamma|\; , \eqlabel{nf}
\end{align}
where $|\Gamma|$ is the order of the freely-acting symmetry $\Gamma$. Note this equation implies that the chiral spectrum always consist of a number of complete GUT families ${\bf 10}\oplus\bar{\bf 5}$. The reason for demanding $3|\Gamma|$ rather than just $3$ such families is that the final model is defined on the quotient Calabi-Yau $\hat{X}=X/\Gamma$ with a bundle $\hat{V}\rightarrow \hat{X}$ which descends from $V$. Since the index satisfies ${\rm ind}(\hat{V})={\rm ind}(V)/|\Gamma|$, Eq.~\eqref{nf} does indeed guarantee three chiral families ``downstairs".  The downstairs model on $\hat{X}$ also allows for the inclusion of an additional flat bundle (a Wilson line) which breaks $SU(5)$ to the standard model gauge group without disturbing the index. In the present paper, we will not be concerned with the downstairs construction. We know that it can be carried out and it is, hence, sufficient to focus on the data of the upstairs GUT model. 

A further physical condition, in addition to Eq.~\eqref{nf}, is
\begin{equation}
 \#\overline{\bf 10}=h^2(X,V)\stackrel{!}{=}0\;, \eqlabel{10bar}
\end{equation}
which ensures the absence of vector-like ${\bf 10}$--$\overline{\bf 10}$ pairs. We also need at least one vector-like $\bar{\bf 5}$--${\bf 5}$ pair to account for the Higgs, which amounts to the condition
\begin{equation}
\#\mbox{pairs of Higgs doublets}=\#{\bf 5}=h^2(X,V)\stackrel{!}{>}0\; . \eqlabel{higgs}
\end{equation}
\vskip 2mm
\noindent In summary, the data $(X,\Gamma, V)$ defines a consistent string compactification, iff the Chern class conditions~\eqref{ccond} and the slope conditions~\eqref{slopecond} are satisfied. The most basic physical requirement on such a model is that is has the correct chiral asymmetry of matter which amounts to the condition~\eqref{nf}. We call a model with these properties, all of which are topological, a standard-like  model (SLM). The next most basic physical requirements are the absence of vector-like pairs of matter, Eq.~\eqref{10bar}, and the presence of multiplets to account for the Higgs, Eq.~\eqref{higgs}. Both of these conditions depend on cohomology and are, hence, not strictly topological.\\[2mm]
For the purpose of machine learning, we need to translate the above geometrical data into numerical, integer data. To this end, we introduce a basis $\{J_i\}$ of the second cohomology and a dual basis $\{\nu^i\}$ of the fourth cohomology of $X$, where $i=1,\ldots ,h$ and $h=h^{1,1}(X)$. Relative to this basis the second Chern class of the tangent bundle can be written as $c_2(TX)=c_{2i}(TX)\nu^i$, with integers $c_{2i}(TX)$. The K\"ahler form of $X$ can be expanded as $J=t^iJ_i$, where $t^i$ are the K\"ahler parameters.\footnote{For simplicity, we assume that the K\"ahler cone of $X$ is given by all $J=t^iJ_i$ with $t^i\geq 0$ and the Mori cone by all positive linear combinations of the $\nu^i$. This is indeed the case for all the CICY three-folds studied in this paper, as listed in~\tabref{data} (see Ref.~\cite{Anderson:2017aux} for the comparison between the K\"ahler cone of a CICY and that of its ambient space).} We will also need the triple intersection numbers of $X$, defined by
\begin{equation}
 d_{ijl}=\int_XJ_i\wedge J_j \wedge J_l\; .
\end{equation} 
Further, a line bundle on $X$ can be represented by an $h$-dimensional integer vector $k=(k^1,\ldots ,k^h)$ and is written as ${\cal O}_X(k)$, such that $c_1({\cal O}_X(k))=k^iJ_i$. This means that the line bundle sum~\eqref{V} takes the form
\begin{equation}
 V=\bigoplus_{a=1}^5{\cal O}_X(k_a)\qquad\leftrightarrow\qquad K=(k_a^i)
\end{equation}
and it can be represented by the $h\times 5$ integer matrix $K$.  In practice, a model, for a fixed Calabi-Yau manifold $X$ and a fixed symmetry $\Gamma$, will be represented by this integer matrix $K$. In terms of this matrix, the Chern classes of such a line bundle sum are given by\footnote{The results for the second Chern class and the index are valid provided that $c_1(V)=0$.}
\begin{eqnarray*}
 c_1(V)&=&\sum\limits_{a=1}^5k^i_aJ_i\;,\\
 c_2(V)&=&-\frac{1}{2}d_{ijl}\sum\limits_ak_a^jk_a^l\nu^i\;,\\
 {\rm ind}(V)&=&\frac{1}{6}d_{ijl}\sum\limits_{a=1}^5k_a^ik_a^jk_a^l\; .
\end{eqnarray*}
Hence, a matrix $K=(k_a^i)$ satisfies the Chern class consistency conditions~\eqref{ccond} iff
\begin{equation}
 \sum_{a=1}^5k_a^i\stackrel{!}{=}0\;,\qquad -\frac{1}{2}d_{ijl}\sum_{a=1}^5k_a^jk_a^l\stackrel{!}{\leq} c_{2i}(TX)\;, \eqlabel{ccond1}
\end{equation}
for all $i=1,\ldots  ,h$ and it satisfies the slope conditions~\eqref{slopecond} iff there exit K\"ahler parameters $t^i> 0$ such that
\begin{equation}
 \mu({\cal O}_X(k_a))=d_{ijl}t^ik_a^jk_a^l\stackrel{!}{=}0 \eqlabel{slopecond1}
\end{equation}
for all $a=1,\ldots ,5$. The condition~\eqref{nf} for the correct chiral asymmetry translates into
\begin{equation}
 \frac{1}{6}d_{ijl}\sum_{a=1}^5k_a^ik_a^jk_a^l\stackrel{!}{=}-3|\Gamma|\; . \eqlabel{nf1}
\end{equation} 
We have now expressed all conditions for a SLM in terms of the matrix $K$. The cohomology conditions~\eqref{10bar} and \eqref{higgs} are not so easily dealt with. Standard methods to compute line bundle cohomology, usually based on a version of Cech cohomology, are complicated and usually algorithmic in nature. On the other hand, it has been observed in Refs.~\cite{Constantin:2018hvl,Klaewer:2018sfl,Larfors:2019sie,Brodie:2019pnz} that line bundle cohomology dimensions can be described by relatively simple formulae, which are piecewise polynomial in the line bundle integers $k^i$. It has also been shown in Ref.~\cite{Brodie:2019dfx} that these formulae can be obtained using machine learning techniques. We will rely on some of these results in \secref{Higgs} when we attempt to machine learn the number of Higgs multiplets.\\[2mm]
%


\section{Data sets}\seclabel{data}

\begin{table}[!!!h]
\begin{center}
\begin{tabular}{|c|c|c|c|c|}\hline
id of $X$&$h=h^{1,1}(X)$&$(c_{2i}(TX))$&$|\Gamma|$&\# SLMs\\\hline\hline
5256&5&$(24, 24, 24, 24, 40)$&4&2128\\\hline
5452&5&$(24, 24, 24, 24, 40)$&4&2122\\\hline
6890&5&$(24, 24, 24, 24, 56)$&2&1750\\\hline
6927&5&$(24, 24, 24, 24, 64)$&4&1264\\\hline
7487&5&$(24, 24, 24, 24, 24)$&4&2115\\\hline
3413&6&$(36, 36, 36, 36, 36, 36)$&3&1737\\\hline
4109&6&$(24, 24, 24, 24, 36, 36)$&2&2058\\\hline
5273&6&$(24, 24, 24, 24, 36, 36)$&2&6753\\\hline
5302&6&$(24, 24, 24, 24, 24, 24)$&2&6294\\\hline
5302&6&$(24, 24, 24, 24, 24, 24)$&4&17329\\\hline
5425&6&$(24, 24, 24, 24, 44, 44)$&2&3128\\\hline
6738&6&$(24, 24, 24, 24, 44, 44)$&2&4243\\\hline
\end{tabular}
\caption{\sf Properties of CICYs with $h^{1,1}(X)\leq 6$ and with at least $1000$ standard-like rank five line bundle models. The id of $X$ refers to the ordering of the standard list in Refs.~\cite{Candelas:1987kf,Green:1986ck,data}.}\tablabel{data}
\end{center}
\end{table}
In Ref.~\cite{Anderson:2013xka}, complete sets of SLMs of the type described above have been found, by brute-force scanning, for all complete-intersection Calabi-Yau manifolds in products of projective spaces (CICYs)~\cite{Candelas:1987kf,Green:1986ck} with $h=h^{1,1}(X)\leq 6$. The data for these manifolds and the lists of matrices $K$ for SLMs on these manifold can be found here~\cite{data}. There are $12$ pairs of $(X,|\Gamma|)$ for which the set of SLMs is sufficiently large (meaning at least $1000$ models) to make machine learning viable and the basic properties for these cases are listed in Table~\ref{tab:data}.
We emphasise that these sets contain models with and without Higgs multiplets and we will use this fact in Section~\ref{sec:Higgs}. As an example, the distribution of the number of Higgs pairs (given by $h^2(X,\wedge^2 V)$) for the largest data set in \tabref{data}, the manifold \#5302 with $|\Gamma|=4$, is shown in \figref{nhiggs}. As is evident from the plot, about a quarter of the SLMs have no Higgs pair and the remaining three quarters have one or more than one pair. 
\begin{figure}[h!!!]
\begin{center}
\includegraphics[width=8cm]{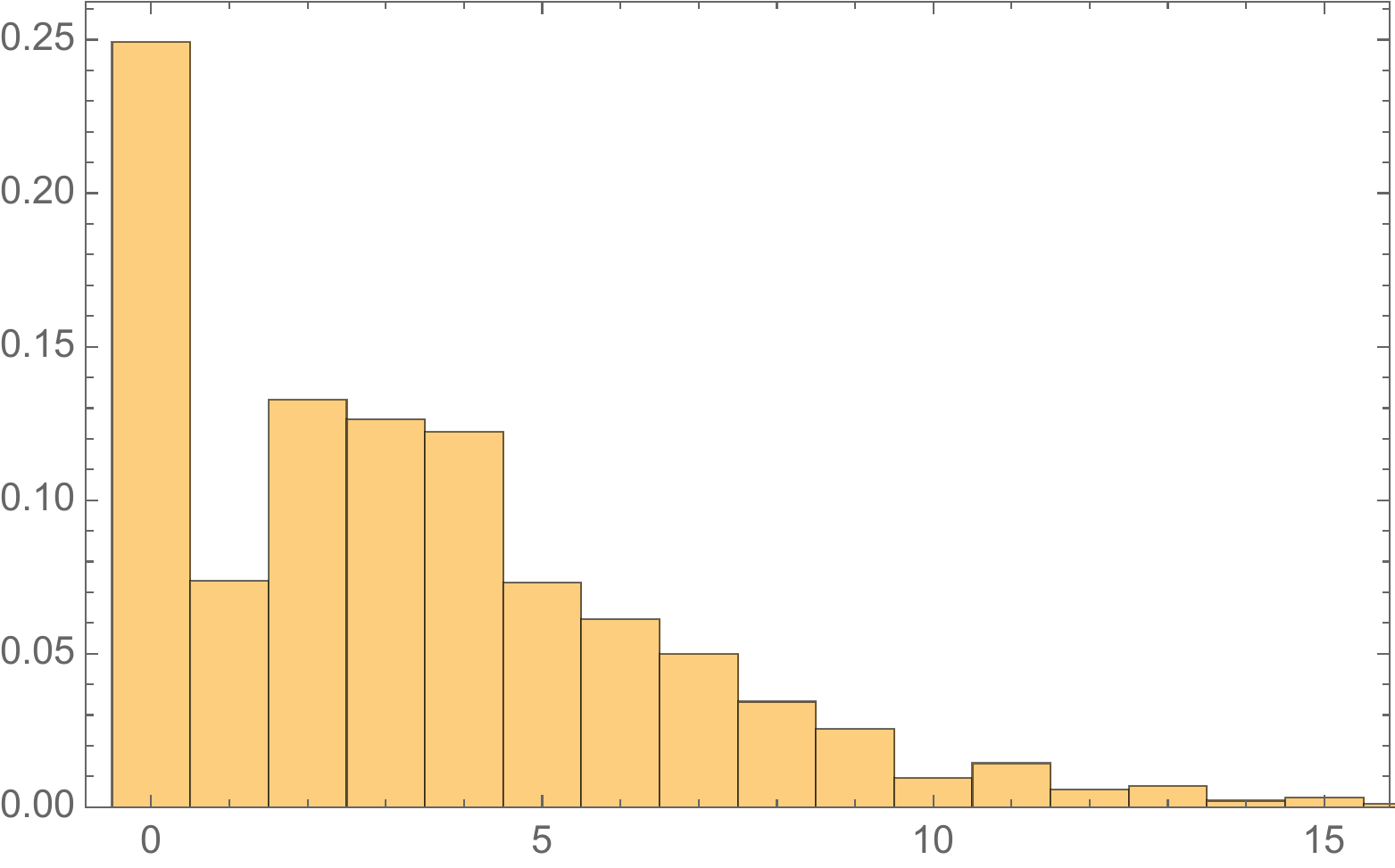}
\caption{\sf  Fractional distribution of the number of Higgs pairs for SLMs on the manifold \#5302 with $|\Gamma|=4$.}\figlabel{nhiggs}
\end{center}
\end{figure}
\newline\noindent
A useful way to characterise models is by the norm
\begin{equation}
 |K|:=\left(\sum_{a,i} |k_a^i|^2\right)^{1/2} \eqlabel{K}
\end{equation} 
which is an indicator of the size of the entries of $K$. As a typical example, the distribution of $|K|$ for CICY \#5302 with $|\Gamma|=4$ is shown in~\figref{datadist} (plot on the left).
\begin{figure*}
\begin{center}
\includegraphics[width=7.5cm]{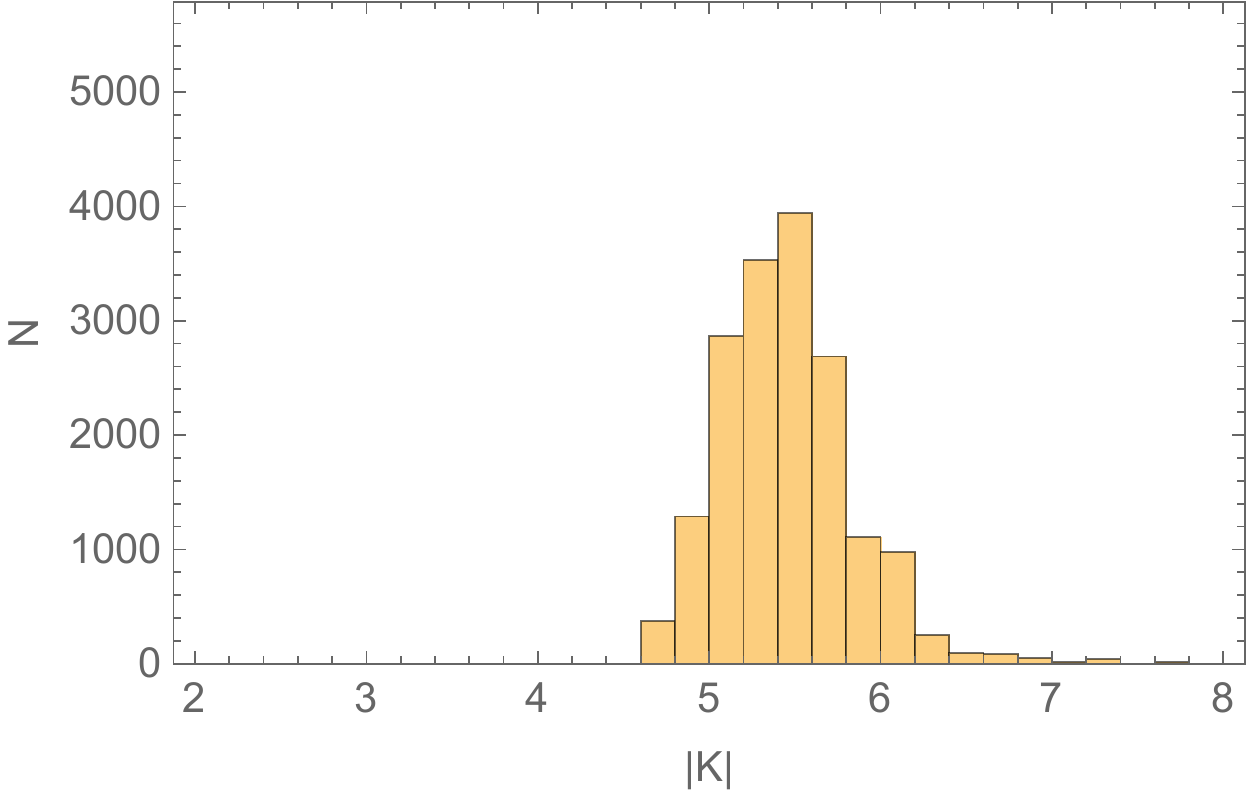}\hskip8mm
\includegraphics[width=7.5cm]{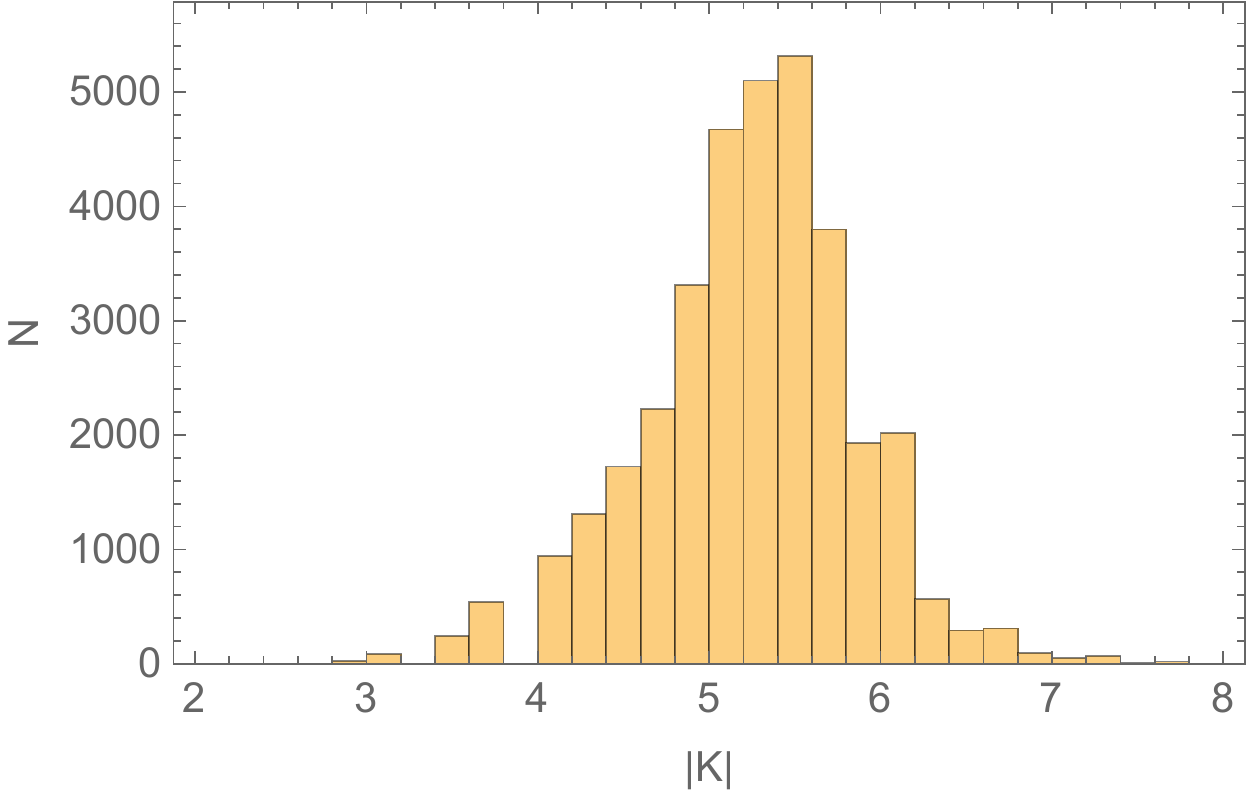}
\caption{\sf  Distribution of $|K|$ in Eq.~\eqref{K} for SLMs (left) and for SLMs and non-SLMs combined (right) for CICY \#5302 and $|\Gamma|=4$. }\figlabel{datadist}
\end{center}
\end{figure*}

For learning of SLMs in \secref{smsimple} and in \secref{smauto} we combine the lists of SLMs with random non-standard models which are defined by random matrices $K$ of the appropriate size. How should these random matrices be generated? One option is to generate random entries in the range observed for the SLMs (typically from $-5$ to $5$) with a flat probability distribution. However, this method generated a distribution of $|K|$-values with a large mean and  practically no overlap with that for the SLMs. Hence, random models generated in this way can be easily distinguished from SLMs by the value of $|K|$.

Instead, we generate entries with a normal distribution, where the mean and width are taken the same as observed for the SLMs. This ensures that the matrices $K$ for SLMs and non SLMs have a similar distribution of entries. For all data sets in \tabref{data}, the mean of the entries is close to $0$ and the standard deviation of the entries is between $1$ and $1.3$, depending on the data set. In addition, we ensure that all random models satisfy the trivial Chern class condition $c_1(V)=0$, that is, the first Eq.~\eqref{ccond1}. For CICY \#5302 and $|\Gamma|=4$ the combined distribution of SLMs and non SLMs is shown on the right-hand side of \figref{datadist}.

What is the distribution of the properties which define SLMs within this set of random non SLMs, generated as described above? The anomaly condition, that is, the second Eq.~\eqref{ccond1}, is satisfied by practically all random models (more precisely, by a fraction of $0.94$--$0.99$ depending on the manifold). This is expected, given we have chosen random matrices with entries similar to those for the SLMs and the condition in question is an inequality. On the other hand, the slope condition~\eqref{slopecond1} is almost never satisfied for the random models (at most for a fraction of $0.01$). Finally, the random models also very rarely satisfy the family condition~\eqref{nf1} (a fraction of $0.02$ to $0.06$, depending on the manifold). A neural network which distinguishes SLMs from random non SLMs will have to be sensitive to at least one of the last two conditions. A network with a success rate close to $1$ needs to be sensitive to each of the conditions separately.\\[2mm]
For the practical machine learning application, we will focus on a particular manifold $X$ and symmetry order $|\Gamma|$, corresponding to one of the rows in \tabref{data}. For supervised learning of the standard model property on this manifold we will use a dataset of the form
\begin{equation} 
 \{K=(k_a^i)\rightarrow 0\mbox{ or } 1\} \eqlabel{Ksm}
\end{equation} 
where $K$ is either a random integer matrix, generated as explained above, describing a non SLM if assigned to $0$ or a matrix describing a SLM if assigned to $1$. For unsupervised learning of the standard model property we use the same dataset $\{K\}$ but with the labels omitted.  

To learn about the Higgs, the datasets will be of the form
\begin{align}
\nonumber
& \{K=(k_a^i)\rightarrow 0\mbox{ or } 1\}\\
& \{K=(k_a^i)\rightarrow\mbox{\# of Higgs pairs}\}\; , \eqlabel{KHiggs}
\end{align} 
where we only include matrices $K$ which describe SLMs. Recall that SLMs can have varying numbers of Higgs pairs, as shown in \figref{nhiggs}. The first above set is appropriate for a simple binary classification of the absence $(0)$ or presence $(1)$ of Higgs multiplets and the second set is used to learn the actual number of Higgs multiplets.

As we will see in \secref{Higgs}, training sets of this form are not quite suited for successful learning of the Higgs property. Instead, we will be using the feature-enhanced data sets
\begin{align}
\nonumber
& \{(k_a^i,k_a^ik_a^j,k_a^i k_a^jk_a^l)\rightarrow 0\mbox{ or } 1\}\\
& \{(k_a^i,k_a^ik_a^j,k_a^i k_a^jk_a^l)\rightarrow\mbox{\# of Higgs pairs}\}\; , \eqlabel{K3Higgs}
\end{align} 
with the quadratic and cubic monomials in the line bundle integers $k_a^i$ added to the input. This method is informed by the observation~\cite{Constantin:2018hvl,Klaewer:2018sfl,Larfors:2019sie,Brodie:2019pnz,Brodie:2019dfx} that line bundle cohomology dimensions on three-folds are described by piecewise cubic formulae.

For the purpose of Higgs learning, we will also find that our datasets as in \tabref{data} are still too small. In this context, it is useful to observe that our models $K=(k_1,\ldots ,k_5)$ have an obvious permutation symmetry $S_5$ which permutes the five line bundles $k_a$. Model properties are of course completely independent of these permutations and the resulting redundancies have already been removed from the datasets in \tabref{data}. Conversely, we can now use these permutation to enhance the size of our datasets. We will return to this point in \secref{Higgs}.\\[2mm]
Throughout the paper, neural networks will be realised both in TensorFlow and in Mathematica, and trained with a standard stochastic gradient descent method and using a mean square loss. Training and validation will be measured by the loss and by the success rate, by which we mean the fraction of models for which the trained network produces the correct integer target after rounding. 

\section{Learning standard models with supervised learning}\seclabel{smsimple}
In this section, we will study whether neural networks can distinguish SLMs from non SLMs. Our datasets are of the form~\eqref{Ksm} and are split into a training set (70\%) and a validation set (30\%). As we will see, for the purposes of this section, it is sufficient to consider relatively simple networks of the form
\begin{equation}
K\in \mathbb{Z}^{5h}\stackrel{L}{\longrightarrow}\mathbb{R}^{16}\stackrel{{\rm selu}}{\longrightarrow}\mathbb{R}^{16}\stackrel{L}{\longrightarrow}\mathbb{R}\stackrel{\sigma}{\longrightarrow}[0,1]\subset \mathbb{R} \eqlabel{nnsm}
\end{equation} 
where $L$ is an affine transformation $x\mapsto Wx+b$ with a weight matrix $W$ and a bias vector $b$ of the appropriate dimensions, selu is the scaled exponential linear unit activation function, defined for each element as 
\beq\eqlabel{selu}
{\rm selu}(x)= \left\{ \begin{array}{l}
x, \mbox{ if } x >0 \\
\alpha e^x - \alpha, \mbox{ if } x \leq 0 \,,
\end{array} \right.
\eeq 
and $\sigma$ is the logistic sigmoid activation function
\beq\eqlabel{ls}
\sigma(x)=(1+e^{-x})^{-1} \,.
\eeq 
We train this network for all cases listed in \tabref{data} and find that the training and validation success rates are always $\geq 0.99$. We conclude that, for a given compactification space, neural networks of the form~\eqref{nnsm} can successfully distinguish SLMs from non SLMs. This works for a range of compactification spaces, as in \tabref{data}.\\[2mm]
The successful generalisation of the network~\eqref{nnsm} for the problem at hand motivates tackling a somewhat more ambitious task. The main problem in searching for SLMs systematically is the shear number of integer matrices $K$ (with $c_1(V)=0$) which increases with a power $4h$. For $h\geq 7$ a full systematic search basically becomes impossible but it is still feasible to scan matrices $K$ with entries constrained in a suitably small range. Suppose we generate a training set of matrices with small entries from such a restricted scan. Does a network trained on such a set generalise to matrices with larger entries?

In order to answer this question, we focus on the manifold \#5302 with $|\Gamma|=4$ from \tabref{data}, which provides our largest data set. We select from this set all matrices with $|K|\leq 5$ (roughly $1/3$ of the models, corresponding to the left part of the distribution in \figref{datadist}), which we split into a training set (70\%) and a validation set (30\%). The other matrices with $|K|>5$ (roughly $2/3$ of the models, corresponding to the right part of the distribution in \figref{datadist}) are used as a test set in order to test the generalisation of the network to matrices with a size beyond the training range. 

For the data structured in this way, we train the network~\eqref{nnsm}. The training and validation loss as a function of training rounds is shown in \figref{train5302} and training and validation success rates are both $\geq 0.99$.
\begin{figure}[h!!!]
\begin{center}
\includegraphics[width=8cm]{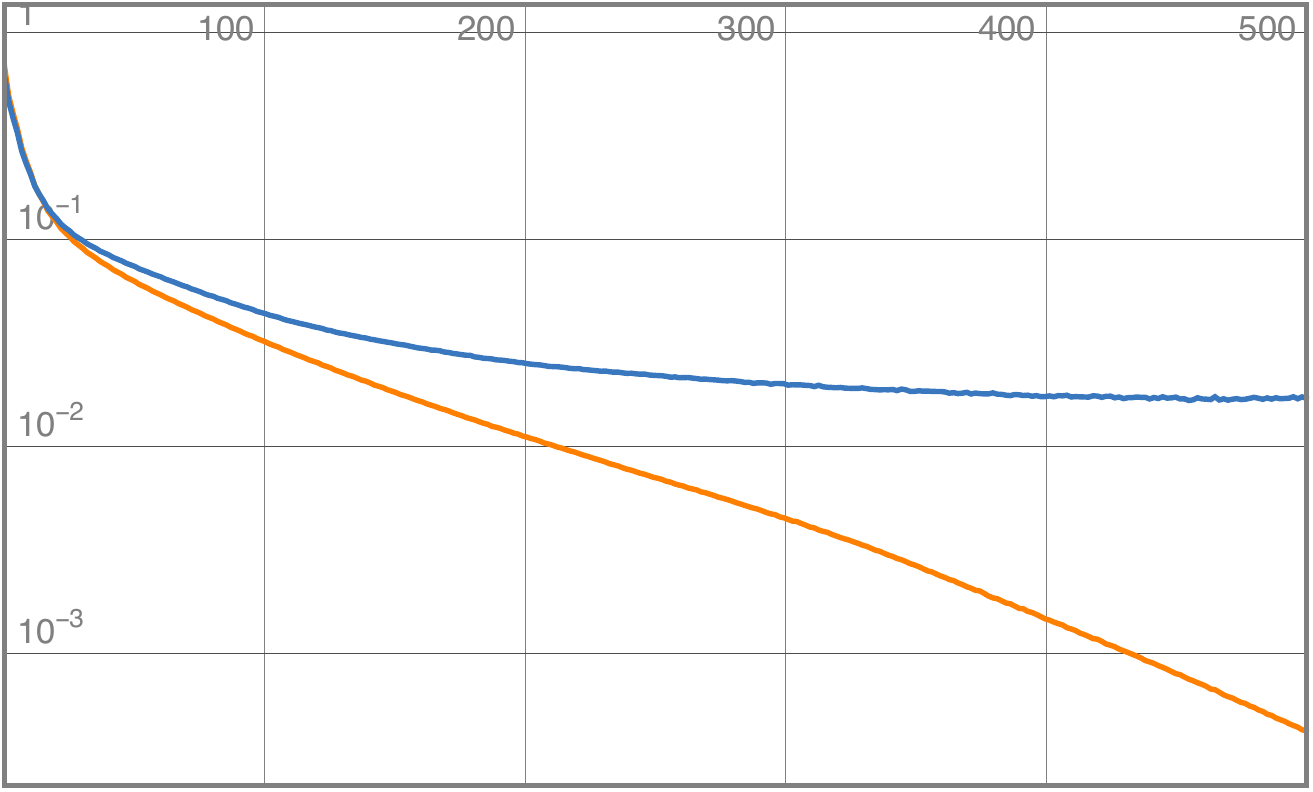}
\caption{\sf  Training loss (orange) and validation loss (blue) for the network~\eqref{nnsm} as a function of training rounds for the data from CICY \#5302 and $|\Gamma|=4$ with $|K|\leq 5$.}\figlabel{train5302}
\end{center}
\end{figure}
Crucially, the success rate of the network trained on this set of matrices with $|K|\leq 5$ on the test set of matrices with $|K|>5$ is $\geq 0.98$
, and the values of Matthew's phi and F-score are $0.991$ and $0.9956$, respectively. 
In other words, the network generalises extremely well to matrices with larger entries.

For this case, it is instructive to look at the three properties which are required for a SLM, namely the anomaly condition (second Eq.~\eqref{ccond1}), the slope condition~\eqref{slopecond1} and the family condition~\eqref{nf1}, separately.  \tabref{confusion} presents the relevant confusion matrix, covering all eight combinations of these three properties, for the complete data set including all values of $|K|$.
\begin{table}[!h]
\begin{center}
\begin{tabular}{|c|c|c|c|}\hline
(anomaly,slope,family)&\# models&non-SMs $(0)$&SMs $(1)$\\\hline\hline
(0, 0, 0)&573&0.96&0.04\\\hline
(0, 0, 1)&33&0.91&0.09\\\hline
(0, 1, 0)&46&0.96&0.04\\\hline
(0, 1, 1)&2&1.00&0.00\\\hline
(1, 0, 0)&15956&0.99&0.01\\\hline
(1, 0, 1)&517&1.00&0.00\\\hline
(1, 1, 0)&197&0.98&0.02\\\hline
(1, 1, 1)&17329&0.02&0.98\\\hline
\end{tabular}
\caption{\sf Confusion matrix for the dataset on the CICY \#5302 with $|\Gamma|=4$ and the network trained on models with $|K|\leq 5$. The left column indicates the various combinations of properties found in the data set and the final two columns give the fractions of those models identified as SLMs and non SLMs.}\tablabel{confusion}
\end{center}
\end{table}
As discussed earlier, the eight combinations of properties have very different frequencies. However, the message from \tabref{confusion} is that the network does well to distinguish SLMs from non SLMs for most combinations of properties. (The combination $(0,1,1)$ of slope zero models with the right family number which fail on the anomaly condition has such a low frequency that the success rates are not conclusive.)

\section{Auto-encoding standard models}\seclabel{smauto}
\begin{figure*}
\begin{center}
\includegraphics[width=7.5cm]{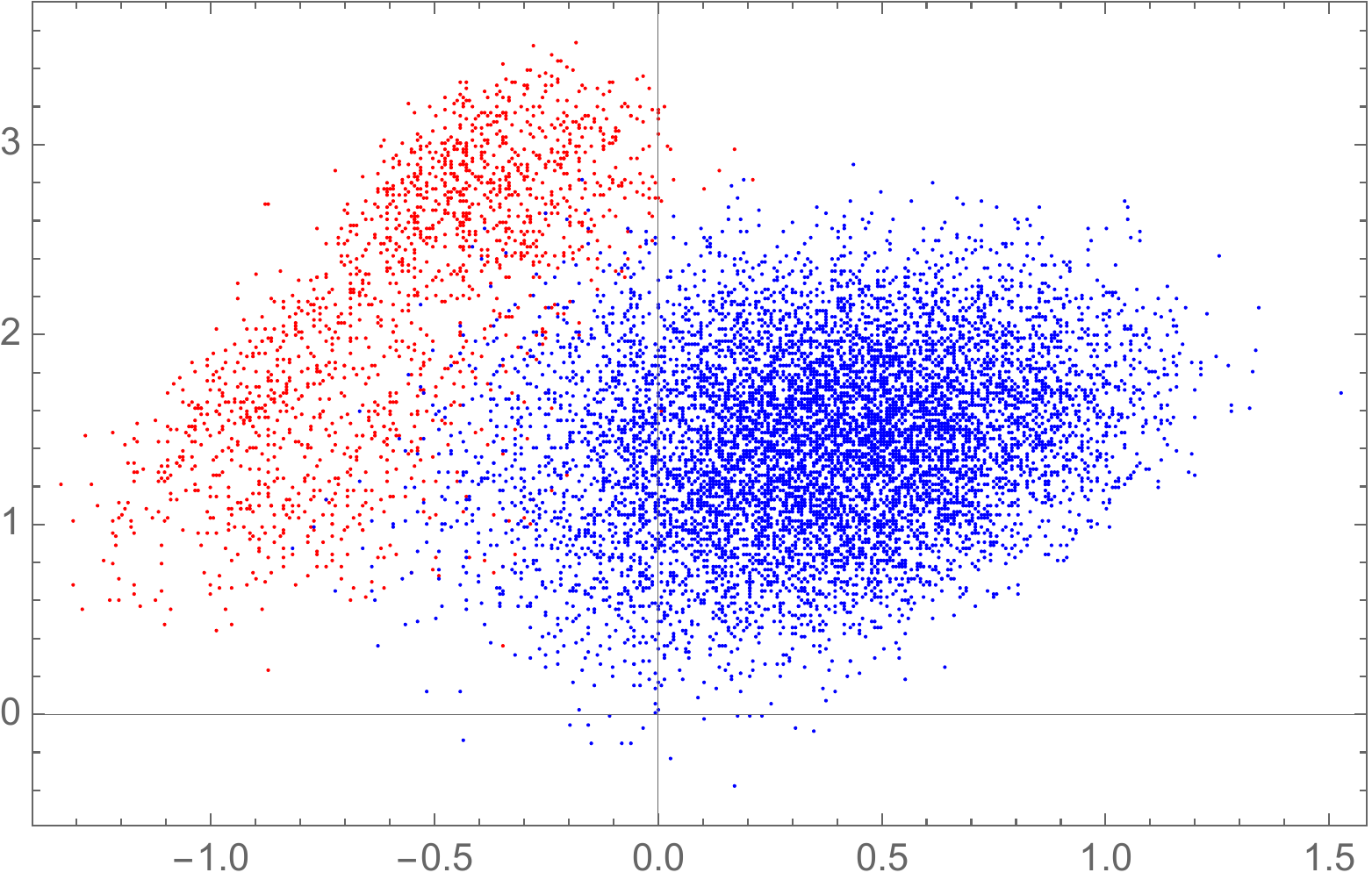}\hskip 8mm
\includegraphics[width=7.5cm]{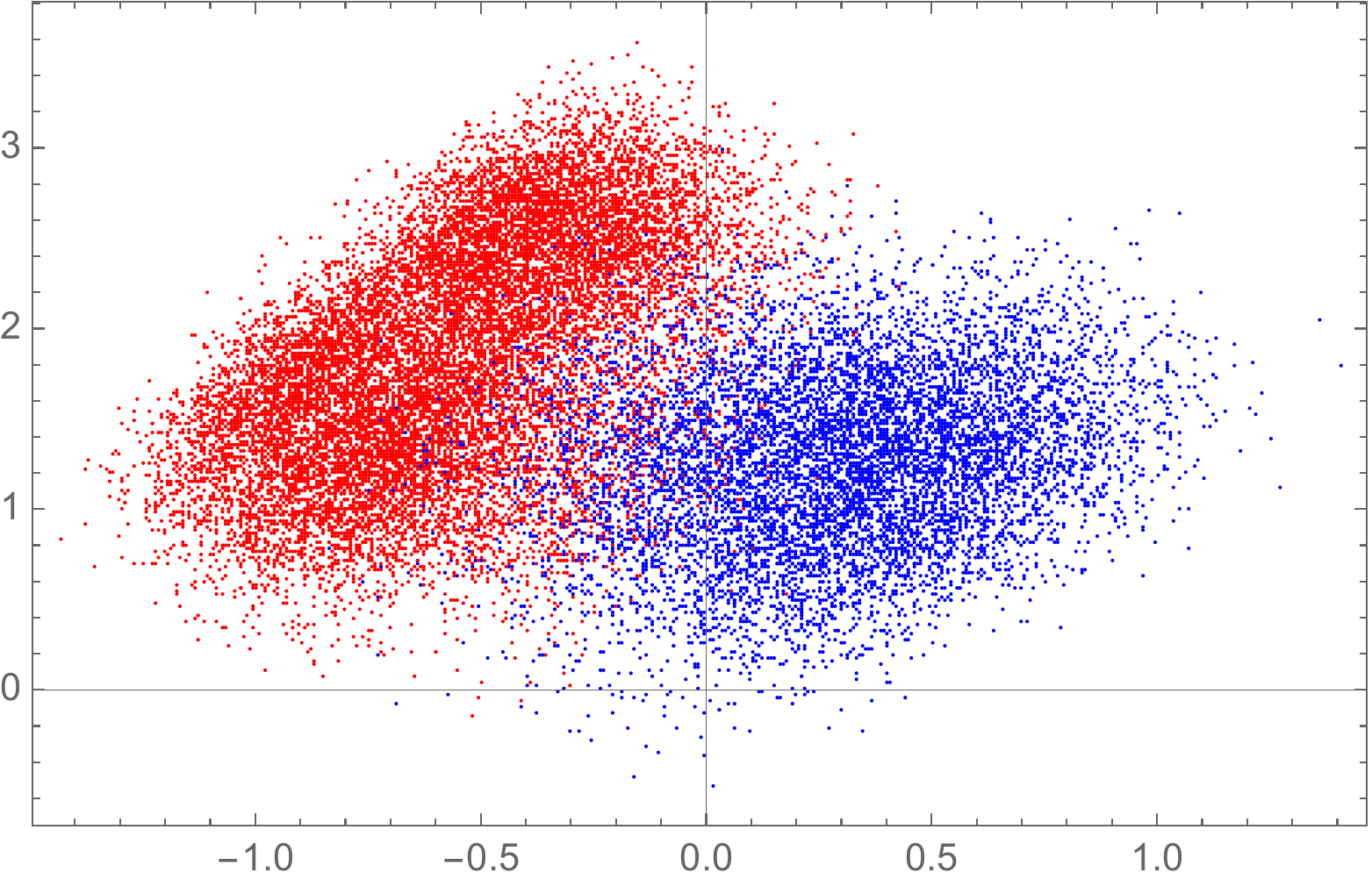}
\caption{\sf  The models on manifold \#5302 with $|\Gamma|=4$  encoded in the two-dimensional latent space of the auto-encoder~\eqref{aeenc}, \eqref{aedec}, trained on models with $|K|\leq 5$. Red points represent SLMs and blue points non SLMs. The plot on the left (right) is for models with $|K|\leq 5$ ($|K|>5$).}\figlabel{ae}
\end{center}
\end{figure*}
We have seen that supervised learning can be used to distinguish SLMs from non SLMs quite efficiently. In this section, we will investigate whether something similar can be accomplished in the context of unsupervised learning, using an auto-encoder.\\[2mm]
We will focus on the manifold \#5302 with $|\Gamma|=4$, our largest training set and, as before, we split into subsets with $|K|\leq 5$, used for training, and with $|K|>5$, used for testing. Guided by the observation in Ref.~\cite{Mutter:2018sra} we also one-hot encode the matrices $K$ which leads to binary input vectors $K_{\rm hot}$ of dimension $330$. The structure of the encoder is
\begin{equation}
K_{\rm hot}\,\in \mathbb{Z}^{330}\stackrel{\tilde{L}}{\longrightarrow}\mathbb{R}^{32}\stackrel{\tilde{L}}{\longrightarrow}\mathbb{R}^{16}\stackrel{\tilde{L}}{\longrightarrow}\mathbb{R}^{8}\stackrel{\tilde{L}}{\longrightarrow}\mathbb{R}^2\; , \eqlabel{aeenc}
\end{equation}
while the decoder
\begin{equation}
\mathbb{R}^2\stackrel{\tilde{L}}{\longrightarrow}\mathbb{R}^{8}\stackrel{\tilde{L}}{\longrightarrow}\mathbb{R}^{16}\stackrel{\tilde{L}}{\longrightarrow}\mathbb{R}^{32}\stackrel{\tilde{L}}{\longrightarrow}\mathbb{Z}^{330}\ni\hat{K}_{\rm hot}\; , \eqlabel{aedec}
\end{equation}
maps between the same dimensions, in reverse order. Here, $\tilde{L}={\rm selu}\circ L$ is the combination of an affine transformation $x\mapsto Wx+b$ between the appropriate dimensions and a selu activation function~\eqref{selu}. Training is performed by minimising the mean square difference $|K_{\rm hot}-\hat{K}_{\rm hot}|$, as usual. Note that the latent space (the output space of the encoder) is $\mathbb{R}^2$, so the compression of the matrix $K$ facilitated by this auto-encoder is to two dimensions.

We remark that one could perform analyses using other methods of dimensional reduction and interestingly arrive at similar results to \figref{ae}.
For instance, a principal component analysis was carried out to reduce the initial labeled data (without hot-encoding but just flattening the bundles into an integer vector of length $6 \times 5 = 30$) directly to 2 dimensions.
One finds that for each of the $|K|\leq 5$ and $|K|>5$ cases, there is a separation between the SLMs and non-SLMs much like \figref{ae}.

From the above auto-encoder, trained with the subset of matrices with $|K|\leq 5$, we then extract the encoder~\eqref{aeenc} and compute the image of all models in the two-dimensional latent space. The results are shown in \figref{ae}.
As is evident from those plots, the separation between SLMs and non SLMs in the two-dimensional latent space is quite convincing.  We emphasise that the auto-encoder was trained on the models with $|K|\leq 5$ only while the models with $|K|>5$ are ``unseen data". Nevertheless, the separation between SLMs and non SLMs occurs for the training set (left plot in \figref{ae}) as well as for the test set (right plot in \figref{ae}). Just as in the case of supervised learning, the auto-encoder, therefore, generalises beyond the training range.

\section{Learning about Higgs multiplets}\seclabel{Higgs}
So far we have used neural networks to distinguish SLMs and non SLMs. The properties separating these two classes of models are topological in nature so we have studied machine learning of topological quantities. In this section, we will be more ambitious and try to learn properties which are non-topological. Specifically, we will restrict our data sets to SLMs and attempt to learn the presence/absence of Higgs multiplets and, in a second step, the number of Higgs multiplets.\\[2mm]
More specifically, we will focus on our largest dataset of SLMs on the manifold \#5302 with $|\Gamma|=4$ with $17329$ models. The distribution of the number of Higgs pairs for this dataset is shown in \figref{nhiggs}. Roughly a quarter of the models have no Higgs pairs and all others have one or more than one Higgs pair. 

We begin with the less ambitious task of learning the presence or absence of Higgs pairs, so the structure of the data set is $\{K\rightarrow 0\mbox{ or }1\}$ as in Eq.~\eqref{KHiggs}. As usual we use 70\% of these models for training and the other 30\% for validation. We consider a variety of networks of the form
\begin{equation}
 K\in\mathbb{Z}^{n}\stackrel{\tilde{L}}{\rightarrow} \mathbb{R}^{n_1}\stackrel{\tilde{L}}{\rightarrow} \cdots \stackrel{\tilde{L}}{\rightarrow} \mathbb{R}^{n_l}\stackrel{L}{\rightarrow}\mathbb{R}\stackrel{\sigma}{\rightarrow}[0,1] \eqlabel{nnhiggs}
 \end{equation}
where $\tilde{L}={\rm selu}\circ L$ combines an affine transformation $L$ with a selu activation~\eqref{selu} and $\sigma$ is a logistic sigmoid activation~\eqref{ls}. We vary the widths $n_i$ as well as the depth $l$ of the network. A typical result for training and validation loss for the choice $l=4$, $n=30$ and $(n_1,n_2,n_3,n_4)=(256,128,64,16)$ is shown in \figref{higgsfail}.
\begin{figure}[h!!!]
\begin{center}
\includegraphics[width=7.9cm]{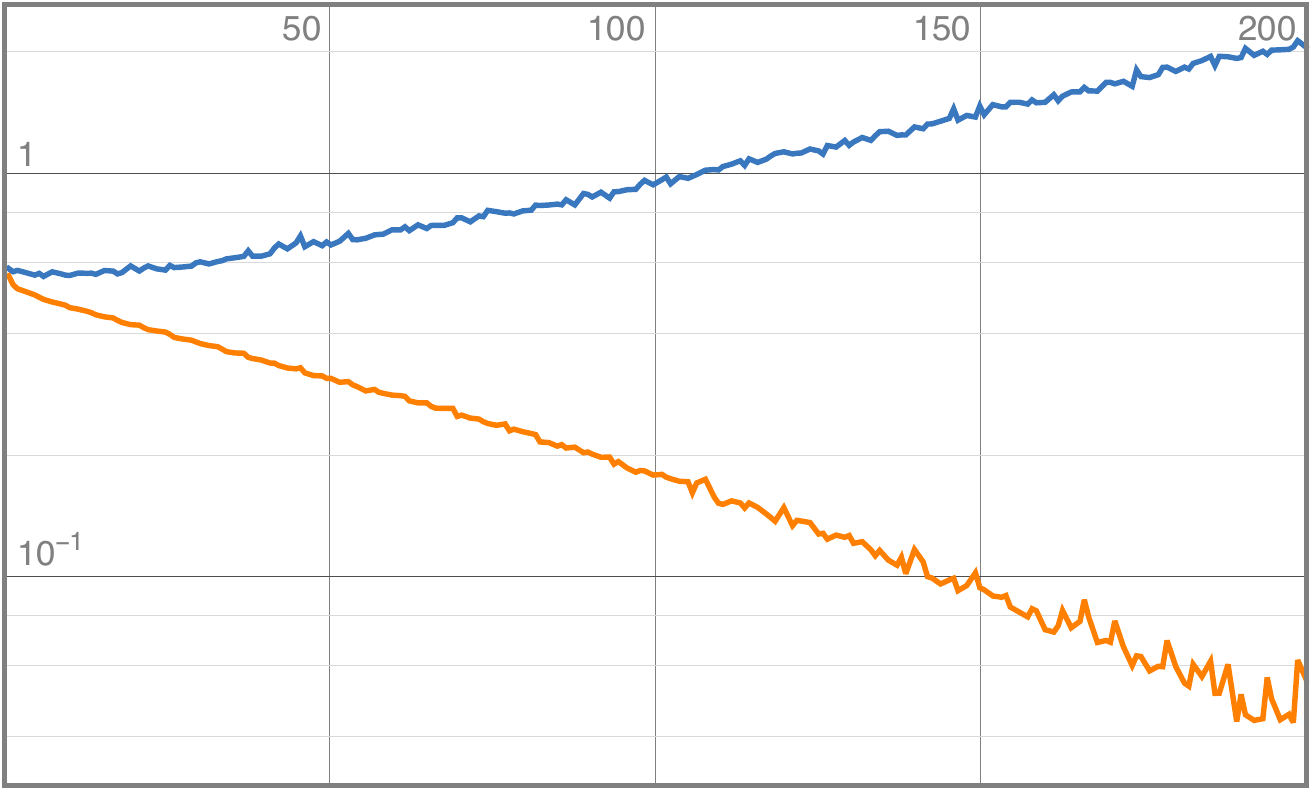}
\caption{\sf  Training loss (orange) and validation loss (blue) for the network~\eqref{nnhiggs} with $l=4$, $n=30$ and $(n_1,n_2,n_3,n_4)=(256,128,64,16)$, attempting to learn the absence/presence of Higgs pairs for SLMs on the manifold \#5032 with $|\Gamma|=4$. The increase of the validation loss indicates a failure to generalise.}\figlabel{higgsfail}
\end{center}
\end{figure}
The confusion matrices for this case are given by
\begin{equation}
\begin{array}{lcc}
&\mbox{training set}&\mbox{validation set}\\[3mm]
&\begin{array}{ll}0\quad&1\end{array}&\begin{array}{ll}0\quad&1\end{array}\\[2mm]
\begin{array}{l}\mbox{target } 0\\\mbox{target }1\end{array}& \left(\begin{array}{cc}0.01&0.23\\0.01&0.75\end{array}\right)&\left(\begin{array}{cc}0.01&0.25\\0.01&0.73\end{array}\right)
\end{array} 
\end{equation} 
Both the loss plot and the confusion matrices indicate complete failure of generalisation. In fact, the confusion matrices shows that all models, including the quarter without Higgs pairs, is identified by the network as having Higgs pairs. We have considered the network~\eqref{nnhiggs} for a variety of depths and widths and with a number of standard measures to improve generalisation, including optimising hyperparameters, different activation functions, inclusion of batch normalisation and dropout layers. The results remain essentially unchanged from the above example. It appears a simple network~\eqref{nnhiggs} is not capable of learning a non-topological property, such as the presence or absence of Higgs pairs.\\[2mm]
What saves the day is an insight into the structure of line bundle cohomology. It is conjectured and, in some cases, proved~\cite{Constantin:2018hvl,Klaewer:2018sfl,Larfors:2019sie,Brodie:2019pnz,Brodie:2019dfx}, that line bundle cohomology dimensions on three-folds are described by piecewise cubic polynomials in the line bundle integers $k^i$. As we have discussed, the number of Higgs pairs is governed by line bundle cohomology, so this insight is relevant. It suggests we should feature-enhance our data sets and add to the line bundle integers $K=(k_a^i)$ their quadrics $(k_a^ik_a^j)$ and their cubics $(k_a^ik_a^jk_a^l)$ (for $i\leq j\leq l$).
This means we now consider a data set of the form $\{(k_a^i,k_a^ik_a^j,k_a^ik_a^jk_a^l)\rightarrow 0\mbox{ or }1\}$, as in Eq.~\eqref{K3Higgs}. 

However, there is one further problem. The dimensions of the enhanced feature space is now $415$ and the available training set of $17329$ models is simply not large enough to train efficiently for such a high dimension. We can solve this problem by using the $S_5$ symmetry which permutes the five line bundles. For each of our models we generate $19$ random permutations and thereby increase the size of our data set by a factor of $20$. As usual, we split this set into a training set (70\%) and a validation set (30\%) and we use a network~\eqref{nnhiggs} with $l=3$, $n=415$ and $(n_1,n_2,n_3)=(256,64,16)$. The result for training and validation loss is in \figref{higgs3}, which shows that both training and validation are successful.
\begin{figure}[h!!!]
\begin{center}
\includegraphics[width=7.9cm]{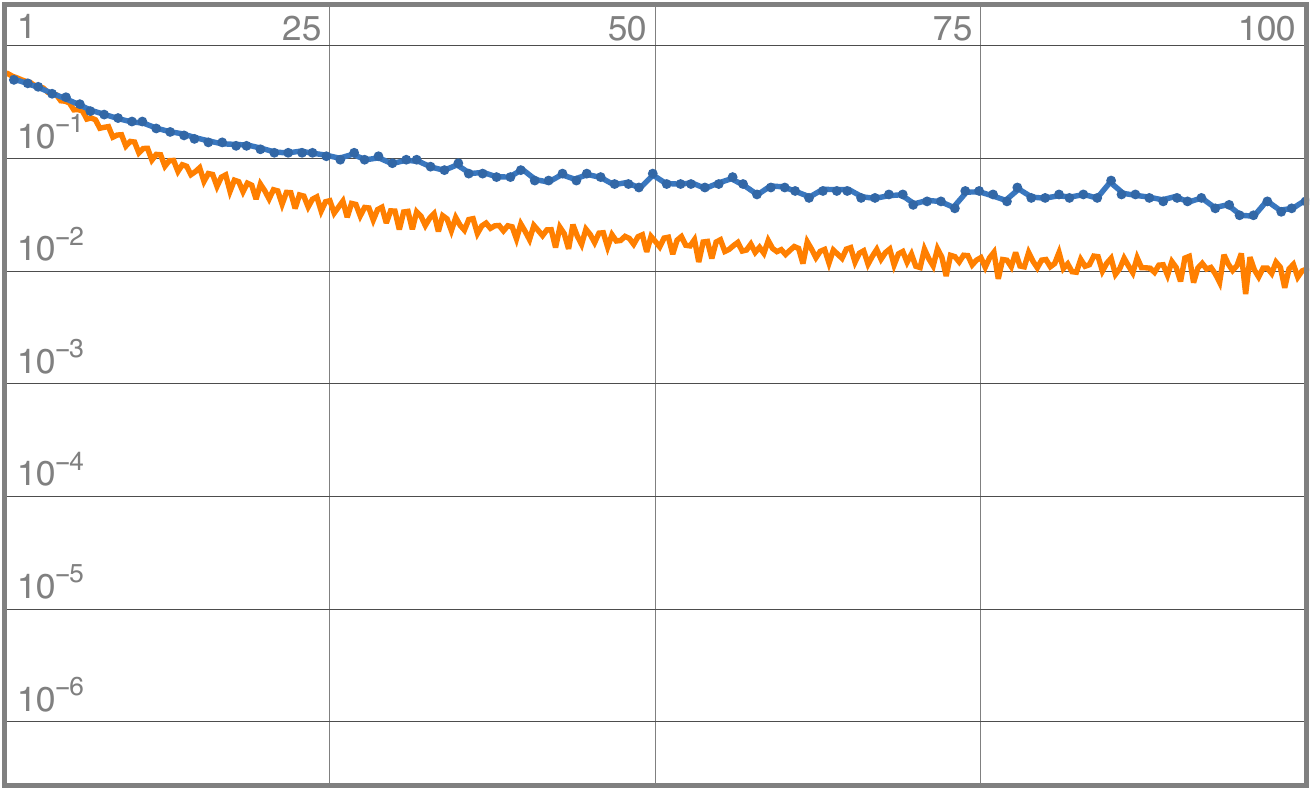}
\caption{\sf  Training loss (orange) and validation loss (blue) for the network~\eqref{nnhiggs} with $l=3$, $n=415$ and $(n_1,n_2,n_3)=(256, 64,16)$, learning the absence/presence of Higgs pairs for SLMs on the manifold \#5032 with $|\Gamma|=4$. A feature-enhanced data set~\eqref{K3Higgs} is used.}\figlabel{higgs3}
\end{center}
\end{figure}
\newline\noindent
The confusion matrix
\begin{equation}
\begin{array}{lcc}
&\mbox{training set}&\mbox{validation set}\\[3mm]
&\begin{array}{ll}0\quad&1\end{array}&\begin{array}{ll}0\quad&1\end{array}\\[2mm]
\begin{array}{l}\mbox{target } 0\\\mbox{target }1\end{array}& \left(\begin{array}{cc}0.25&0\\0&0.75\end{array}\right)&\left(\begin{array}{cc}0.25&0\\0&0.75\end{array}\right)
\end{array} 
\end{equation} 
indicates models with and without Higgs pairs are recognised with success rate $1$.\\[2mm]
Encouraged by this success we now attempt to learn the number of Higgs multiplets for manifold \#5302 with $|\Gamma|=4$, using a feature-enhanced data set $\{(k_a^i,k_a^ik_a^j,k_a^ik_a^jk_a^l)\rightarrow \mbox{\# Higgs pairs}\}$, as in Eq.~\eqref{K3Higgs}. As before, we increase the size of the data set $20$-fold by using the invariance under permutations of the five line bundles. Networks are similar to \eqref{nnhiggs} but with the final logistic sigmoid activation omitted, as appropriate for more general integer target values:
\begin{equation}
 K\in\mathbb{Z}^{n}\stackrel{\tilde{L}}{\longrightarrow} \mathbb{R}^{n_1}\stackrel{\tilde{L}}{\longrightarrow} \cdots \stackrel{\tilde{L}}{\longrightarrow} \mathbb{R}^{n_l}\stackrel{L}{\longrightarrow}\mathbb{R} \eqlabel{nnhiggsnum}\; .
 \end{equation}
 A network of this type with depth $l=3$, $n=415$ and $(n_1,n_2,n_3)=(256,64,16)$ trains and validates successfully, as shown in \figref{higgs3num}.
\begin{figure}[h!!!]
\begin{center}
\includegraphics[width=7.9cm]{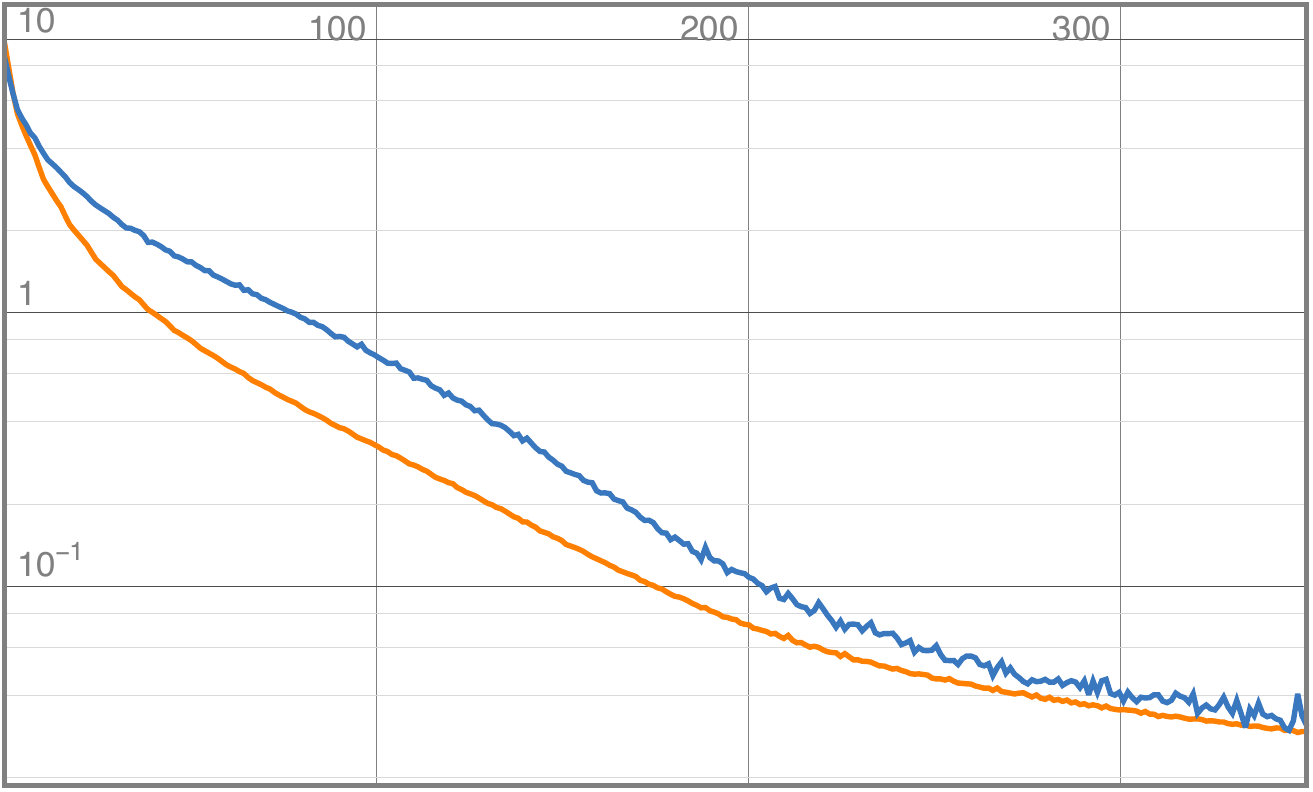}
\caption{\sf  Training loss (orange) and validation loss (blue) for the network~\eqref{nnhiggsnum} with $l=3$, $n=415$ and $(n_1,n_2,n_3)=(256,64,16)$, learning the number of Higgs pairs for SLMs on the manifold \#5032 with $|\Gamma|=4$. A feature-enhanced data set~\eqref{K3Higgs} is used.}\figlabel{higgs3num}
\end{center}
\end{figure}
The confusion matrices for both training and validation sets are nearly diagonal, with all off-diagonal entries $<0.01$, and the diagonal entries correctly reproduce the fractional distribution in \figref{nhiggs}, with deviations uniformly less than $0.01$.

\section{Conclusion}\seclabel{conclusion}
In this paper, we have demonstrated that neural networks are capable of learning phenomenologically relevant properties of string compactifications. Our data sets have been constructed from heterotic line bundle models~\cite{Anderson:2011ns,Anderson:2012yf,Anderson:2013xka}, one of the few data sets where considerable numbers of models with the standard model spectrum have been constructed. 

For a given Calabi-Yau three-fold $X$, a line bundle model is defined by a sum $V=\oplus_{a=1}^5{\cal O}_X(k_a)$ of five line bundles, where $k_a$ are $h=h^{1,1}(X)$ dimensional integer vectors classifying line bundles. Hence, such models can be represented by a $h\times 5$ integer matrices $K=(k_a^i)$ which constitute the features in our applications to machine learning. In Ref.~\cite{Anderson:2013xka} complete intersection Calabi-Yau manifolds with $h\leq 6$ have been scanned for phenomenologically interesting line bundle models. Sizeable data sets (see \tabref{data}) of standard-like models (SLMs), that is consistent models with the correct gauge group and the correct chiral asymmetry, have been found. These data sets, described in \tabref{data}, form the basis of the machine learning applications we have discussed. Many further properties of these models have been computed~\cite{Anderson:2012yf} and we will rely on the results for the number of Higgs pairs.

Our first task has been to learn topological properties of line bundle models. To this end, we have merged SLMs for a given Calabi-Yau manifold with similar sized data sets of random models to obtain sets of the form $\{K\rightarrow 0\mbox{ or } 1\}$, where $1$ indicates an SLM and $0$ a random model. Training is carried out one manifold at a time. Relatively simple networks of the form~\eqref{nnsm} lead to near perfect success rates of $\geq 0.99$ on both training and validation sets for all cases in \tabref{data}. In a more ambitious approach, for the manifold $\#5302$ with symmetry order $|\Gamma|=4$ (see \tabref{data}), we have used a training set drawn from matrices $K$ with small entries ($|K|\leq 5$) and a test set of matrices with large entries ($|K|>5$). It turns out that neural networks~\eqref{nnsm} trained on such a training set still perform extremely well, at a success rate of $\geq 0.98$, on the test set. In other words, the neural network generalises well beyond the ``training range".

For the same data set, on manifold \#5302, split into a training set with small entries and a test set with large entries, we have considered unsupervised learning with an auto-encoder~\eqref{aeenc}, \eqref{aedec}. The results, summarised in \figref{ae}, show that the auto-encoder is capable of distinguishing SLMs from non SLMs and that it retains this capacity on the test set. In particular, this means the generalisation beyond the training range observed for supervised learning, persists for the auto-encoder. 

The final task has been to learn a non-topological property, namely the absence or presence (or, more ambitiously, the number) of Higgs pairs. This property is determined by the values of line bundle cohomology dimensions. For this task we have focused on the data set of SLMs on manifold \#5302 with symmetry order $|\Gamma|=4$ which provides the largest number of models (about $17000$). We have started with simple data sets $\{K\rightarrow 0\mbox{ or } 1\}$, where $1$ is for the presence and $0$ for the absence of Higgs pairs, and fully connected networks of the form~\eqref{nnhiggs}. This simple approach resulted in complete failure to generalise, as exemplified in \figref{higgsfail}. Inspired by the observation~\cite{Constantin:2018hvl,Klaewer:2018sfl,Larfors:2019sie,Brodie:2019pnz,Brodie:2019dfx} that line bundle cohomology dimensions on three-folds are described by piecewise cubic expressions in the line bundle integers $k_a^i$ we have feature-enhanced our data set to $\{(k_a^i,k_a^ik_a^j,k_a^ik_a^j k_a^l)\rightarrow 0\mbox{ or }1\}$, that is, by including quadratic and cubic monomials. Given the significantly increased dimension of the feature space (to $415$) we have also increased the size of the data set by using the invariance under permutations of the five line bundles. A simple fully connected neural network~\eqref{nnhiggs} trained with this enhanced data set does indeed succeed and distinguishes models with and without Higgs pairs reliably. We have also trained a network~\eqref{nnhiggsnum} with a data set  $\{(k_a^i,k_a^ik_a^j,k_a^ik_a^j k_a^l)\rightarrow \mbox{\# of Higgs pairs}\}$ to predict the number of Higgs multiplets successfully.

Our results show that neural networks are capable of learning both topological and non-topological properties of string vacua and are able to select models with phenomenological promise. Recognising non-topological properties is considerably more difficult and has been achieved by feature engineering based on insight into the underlying mathematical structure. The observed generalisation beyond the training range is encouraging. It implies that networks trained on models with small flux integers - which may be obtained by systematic scanning - can be used to identify promising models for larger flux integers. Our results may also be helpful in designing more complicated networks, required for generative approaches (GANs) or reinforcement learning, for instance.

It would be interesting to study whether the present results for heterotic line bundle models persist for other classes of string models, for example for F-theory models or for heterotic models with non-Abelian bundles. 
 
\section*{Acknowledgments}
RD is supported by a Skynner Fellowship of Balliol College, Oxford.
YHH acknowledges STFFC UK, for grant ST/J00037X/1. The work of SJL is supported by IBS under the project code, IBS-R018-D1. AL gratefully acknowledges support from the Simons Center for Geometry and Physics, Stony Brook University at which some of the research for this paper was performed.


\end{document}